\documentclass[%
 reprint,
superscriptaddress,
 amsmath,amssymb,
 aps,
pre,
]{revtex4-2}
\usepackage[usenames,dvipsnames]{xcolor}
\usepackage{hyperref}
\hypersetup{
colorlinks=true,
citecolor=NavyBlue,
linkcolor=NavyBlue,
filecolor=NavyBlue,
urlcolor=NavyBlue
}

\usepackage{comment}
\usepackage{graphicx}
\usepackage{dcolumn}
\usepackage{bm}
\usepackage{orcidlink}
\usepackage[normalem]{ulem}
\usepackage{xcolor}

\begin{document}

\title{Physics-Informed Neural Network for Elastic Wave-Mode Separation} 
\author{E. A. B. Alves}
\author{P. D. S. de Lima}
\email{paulo.douglas.lima@fisica.ufrn.br}
\author{D. H. G. Duarte}
\affiliation{Departamento de Física Teórica e Experimental, \\Universidade Federal do Rio Grande do Norte, 59078-970 Natal-RN, Brazil}
\author{M. S. Ferreira}
\affiliation{School of Physics, Trinity College Dublin, Dublin 2, Ireland}%
\affiliation{Centre for Research on Adaptive Nanostructures and Nanodevices (CRANN) \& Advanced Materials and Bioengineering Research (AMBER) Centre, Trinity College Dublin, Dublin 2, Ireland}
\author{J. M. de Araújo}
\author{C. G. Bezerra}
\affiliation{Departamento de Física Teórica e Experimental, \\Universidade Federal do Rio Grande do Norte, 59078-970 Natal-RN, Brazil}

\date{\today}

\begin{abstract}
Mode conversion in non-homogeneous elastic media makes it challenging to interpret physical properties accurately. Decomposing these modes correctly is crucial across various scientific areas. Recent machine learning approaches have been proposed to address this problem, utilizing the Helmholtz decomposition technique. In this paper, we investigate the capabilities of a physics-informed neural network (PINN) in separating P and S modes by solving a scalar Poisson equation. This scalar formulation offers a dimensionally scalable reduction in computational cost compared to the traditional vector formulation. We verify the proposed method in both homogeneous and realistic non-homogeneous elastic models as showcases. The obtained separated modes closely match those from conventional numerical techniques, while exhibiting reduced transverse wave leakage.
\end{abstract}
\maketitle

\section{Introduction}

Understanding the dynamics of elastic waves in solids is crucial for a wide range of applications, including material characterization~\cite{intro1, intro2, intro3, intro4} and seismology~\cite{Pujol_2003, Tromp2020}. These waves can be described by two modes, corresponding to longitudinal P-waves and transverse S-waves. However, when either P or S waves encounter a material discontinuity, the scattering process generates both P and S modes, a phenomenon known as mode conversion. While metamaterials can be engineered to control or suppress mode conversion fully~\cite{meta1, meta2, meta3, meta4, meta5}, it remains unavoidable in many other cases. Examples abound in the manufacturing, motoring, aerospace, and petroleum industries, to name but a few. In the latter case, this is particularly interesting in the field of seismology, where it presents significant challenges in seismic imaging. Thus, mode separation is crucial for correctly interpreting physical properties in realistic complex media.

The Helmholtz decomposition can separate P and S modes in isotropic and anisotropic~\cite{FANG20241597} media by imposing a divergence-free (for S wave) and a curl-free (for P wave) condition~\cite{morse1953methods}. In the context of multiparameter inverse problems, this decomposition can mitigate ambiguities caused by the fact that different observational signatures are indistinguishable, a phenomenon known as crosstalk~\cite{tromp2005}. Therefore, suppressing this effect can enhance specific seismic imaging methods~\cite{cross1, cross2, cross3, cross5, cross4}. 

Wavefield separation can be efficiently computed by solving an auxiliary vector Poisson equation, where the source term corresponds to the elastic displacement~\cite{Zhu}. The computational cost can be substantially reduced by transforming it into a scalar equation~\cite{ZhengYao}. In both vector and scalar approaches, P and S waves are subsequently constructed by employing adequate vector operations on the solution of the Poisson equation. Therefore, regardless of the separation method used, it typically involves solving a partial differential equation (PDE).

In this regard, physics-informed neural networks (PINNs) have emerged as a practical machine learning approach for solving PDEs~\cite{RAISSI2019686, Karniadakis2021}. PINN belongs to a class of neural networks that incorporates prior knowledge of physical laws into the training process. Such information enables PINN to obtain a superior solution compared to standard data-driven techniques without requiring large amounts of data.

Besides its application in different areas~\cite{PINN1, ZHONG2022133430, DL&INVDIS, PhysRevD.106.124047, BOUSSINESQ, PhysRevLett.132.010801}, PINN has been successfully employed in wave dynamics description~\cite{ALKHALIFAH202111, Song_Chao_Alkha, ORPINN} and inverse-problem solutions~\cite{HAN2025134362, NEGAHDARI2025134518}. In particular, some authors have addressed the mode decomposition problem using various deep learning techniques~\cite{deep1, deep3, deep2, deep4}. Recently, a physical-constrained neural network~\cite{PCNN} and a particular PINN architecture~\cite{mu2025separationpinnphysicsinformedneuralnetworks} have been proposed to separate the elastic wavefield following the Helmholtz decomposition. However, they solve a vector Poisson equation, which leads to increased neural network complexity as the number of spatial dimensions increases. Conversely, the scalar framework can reduce this cost by a factor of $1/2$ in 2D and $1/3$ in 3D problems~\cite{ZhengYao}. Therefore, this formulation offers a scalable computational-cost reduction compared to the traditional vector approach, enabling the use of a less complex neural network architecture.

With that motivation, we investigate PINN capabilities in separate P/S elastic modes by solving a scalar Poisson equation. Furthermore, although the focus of the present communication is in the seismic context, our findings are transferable to other systems containing elastic waves in solid media. This paper is organized as follows. We revisit the elastic wavefield separation in sec.\ \ref{sec:elastic_sep}, highlighting the differences between the vector and scalar formulation. sec.\ \ref{sec:pinn} describes the Fourier feature-embedded PINN architecture used for solving the Poisson equation. The decomposed modes for a simple homogeneous and a realistic non-homogeneous elastic model are presented in Section \ref{sec:results}, followed by conclusions in Section \ref{conc}.

\begin{figure}[!htpb]
    \centering    
    \includegraphics[width=\columnwidth]{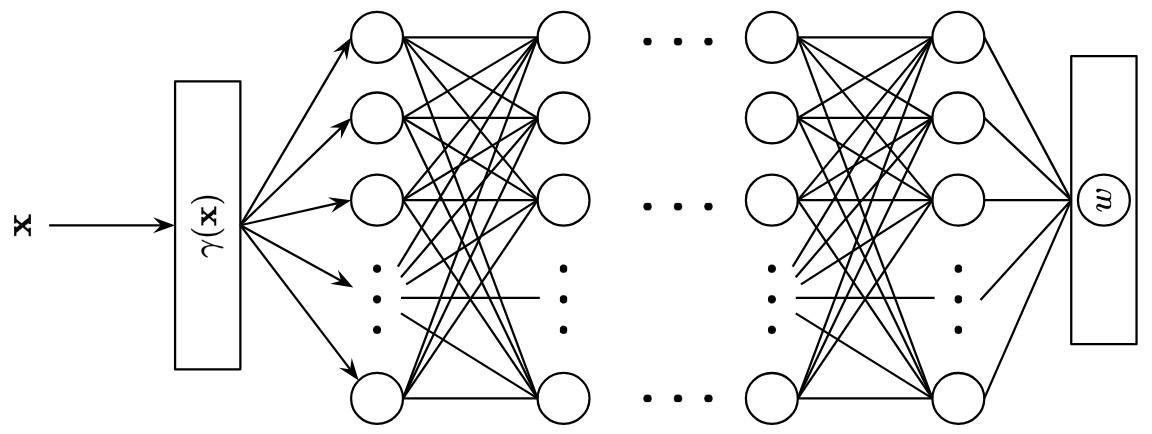}
    \caption{The architecture of the Fourier features physics-informed neural network used to separate the elastic wavefield by solving Eq.\ \eqref{eq:scalar_poisson}, which only requires one output $w(\mathbf{x})$.}
    \label{fig:pinn}
\end{figure}

\begin{figure}[!htpb]
    \centering
    \includegraphics[width=\columnwidth]{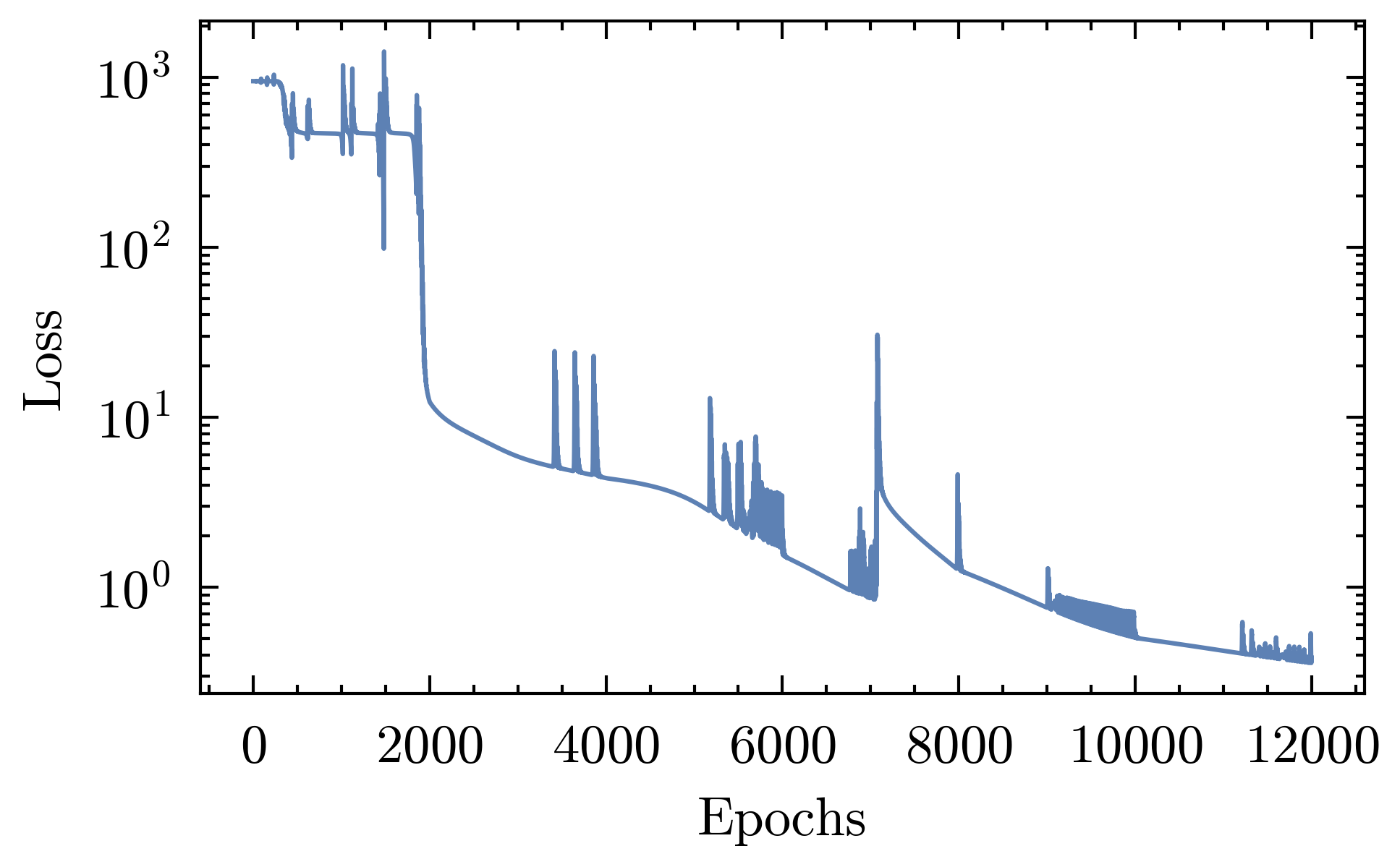}
    \caption{The loss history for the homogeneous model. The learning rate was halved every 2000 epochs.}
    \label{fig:Total_loss_vp}
\end{figure}

\begin{figure*}[!htpb]
    \centering
    \includegraphics[width=0.5\textwidth]{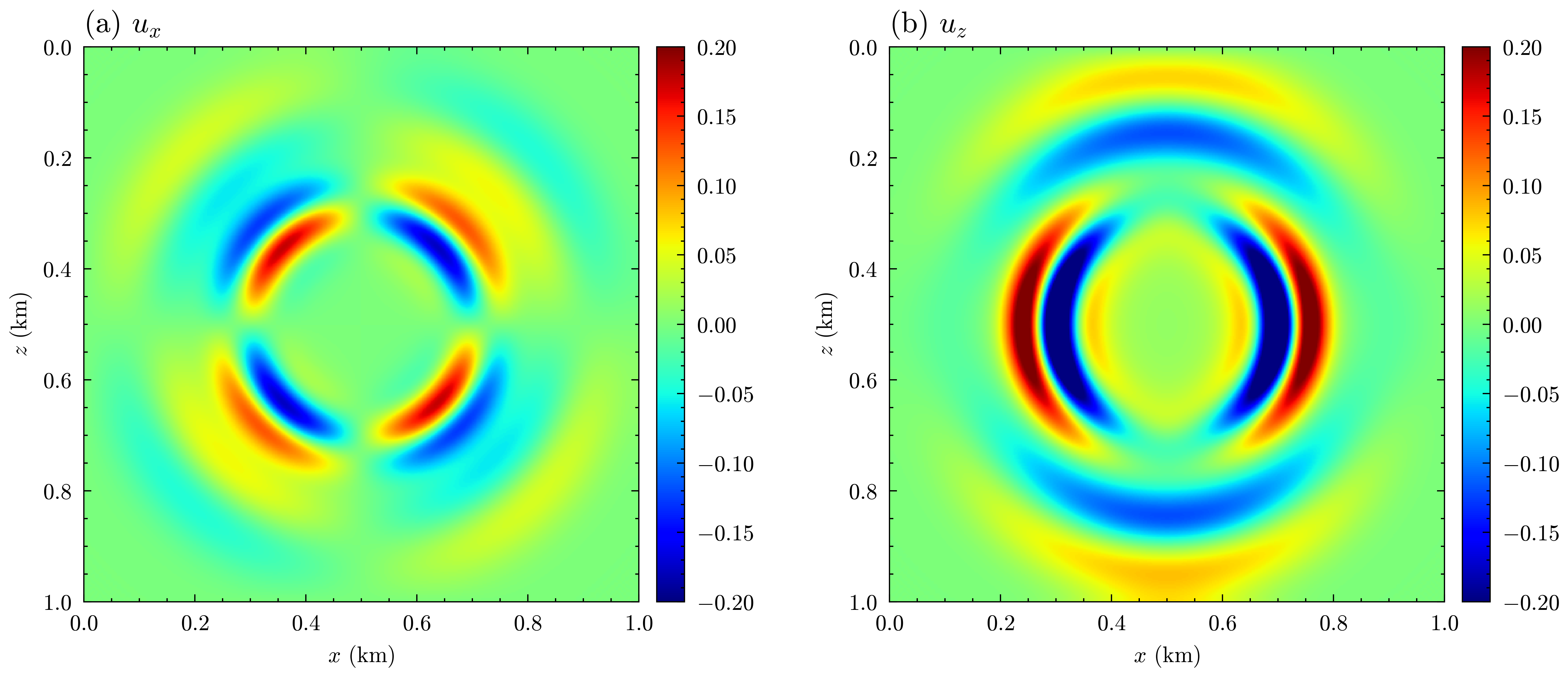}
    \includegraphics[width=\textwidth]{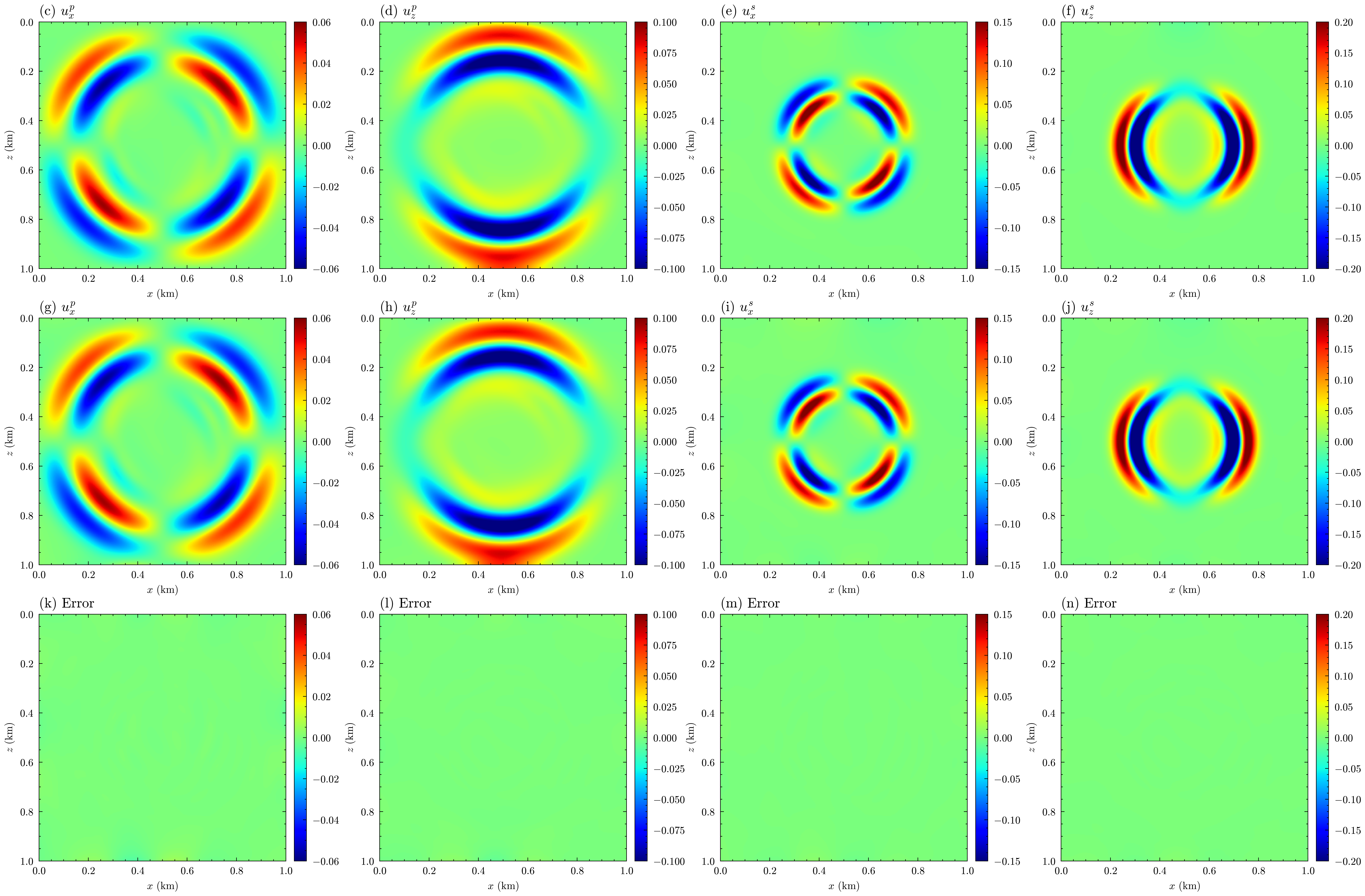}
    \caption{The first two panels show the horizontal and vertical components of the elastic wavefield in a homogeneous medium for a specific instant of time. The separated elastic modes obtained with the conventional numerical solver (second row) and the proposed PINN approach (third row). The difference between them (fourth row) illustrates that both approaches yield very similar results.}
    \label{fig:u_dec_homo_iso_comparation}
\end{figure*}

\begin{figure}[!htpb]
    \centering
    \includegraphics[width=\columnwidth]{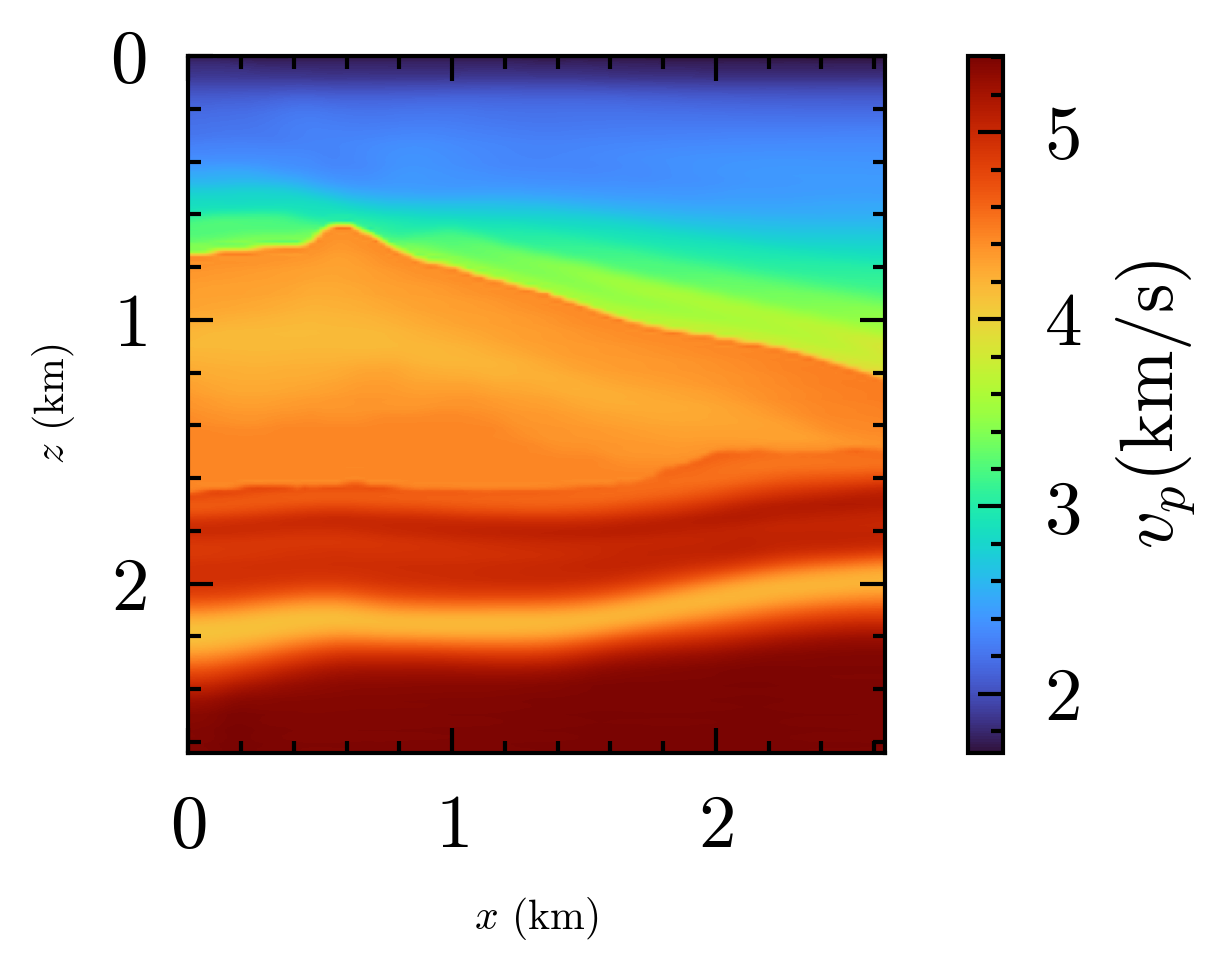}
    \caption{A typical Brazilian pre-salt P velocity model used as a realistic example in the separation mode.}
    \label{fig:GDM_velocity_model}
\end{figure}


\begin{figure}[!htpb]
    \centering
    \includegraphics[width=\columnwidth]{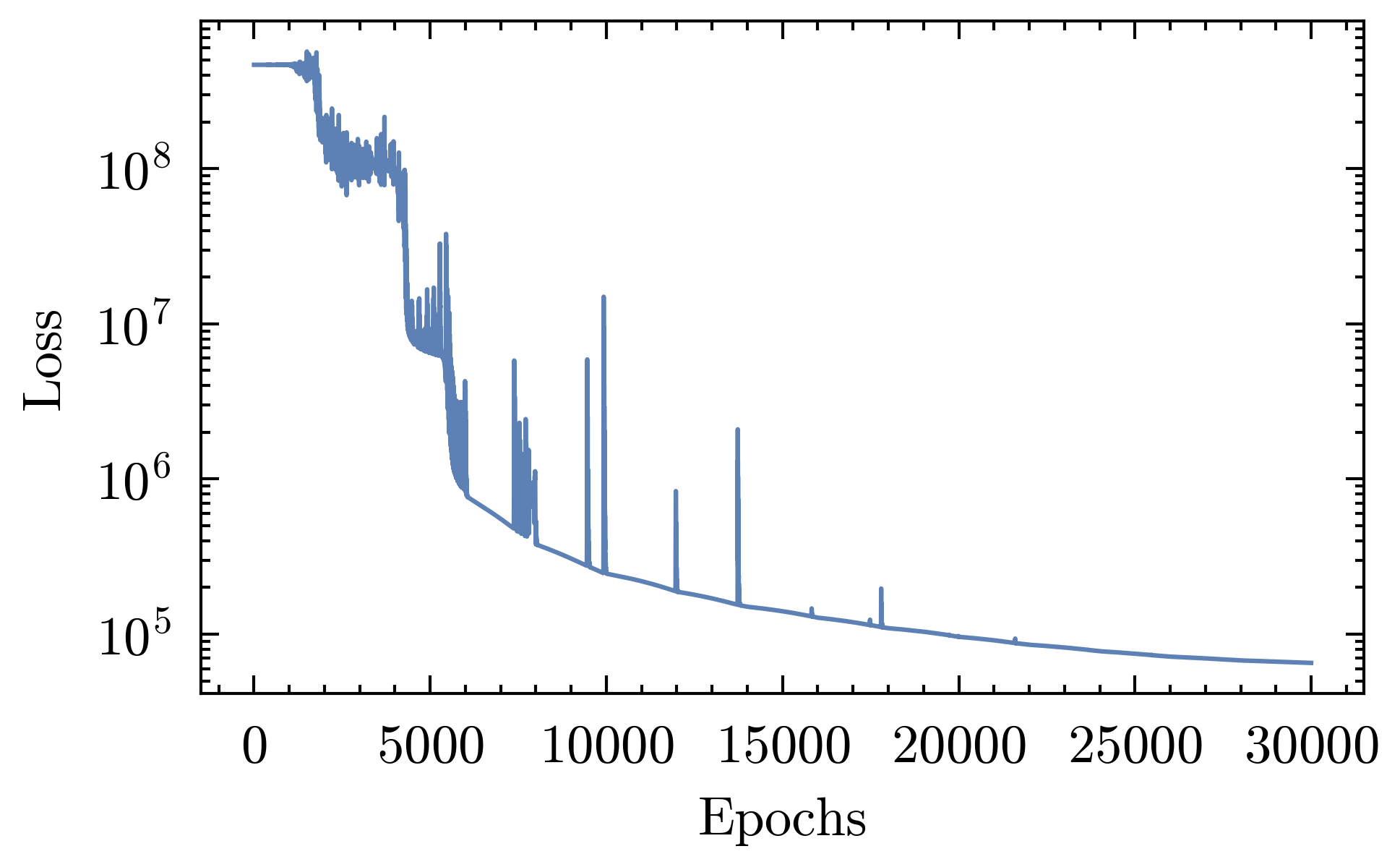}
    \caption{The same as Fig.\ \ref{fig:Total_loss_vp}, but for the non-homogeneous velocity model Fig.\ \ref{fig:GDM_velocity_model}.}
    \label{fig:Total_loss_gdm}
\end{figure}

\begin{figure*}[!htpb]
    \centering
    \includegraphics[width=0.5\textwidth]{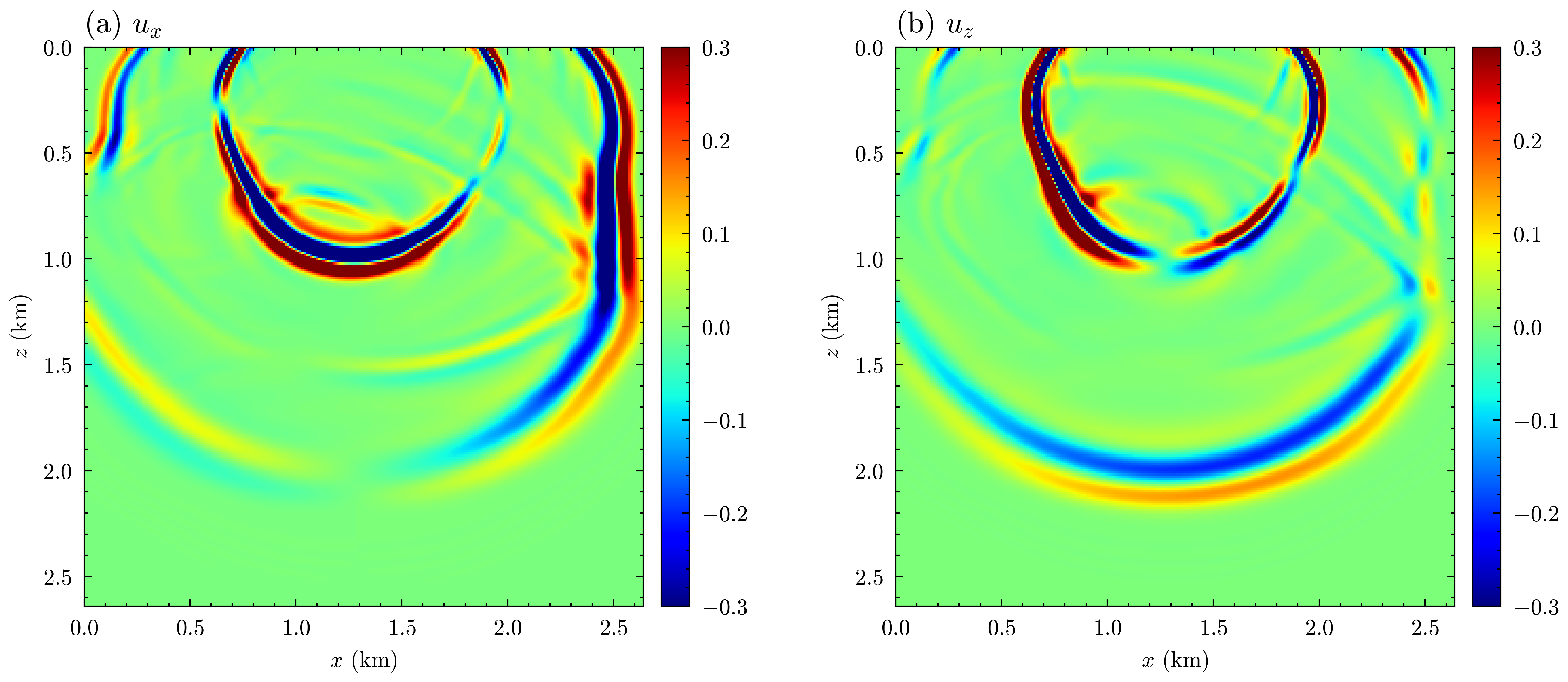}
    \includegraphics[width=\textwidth]{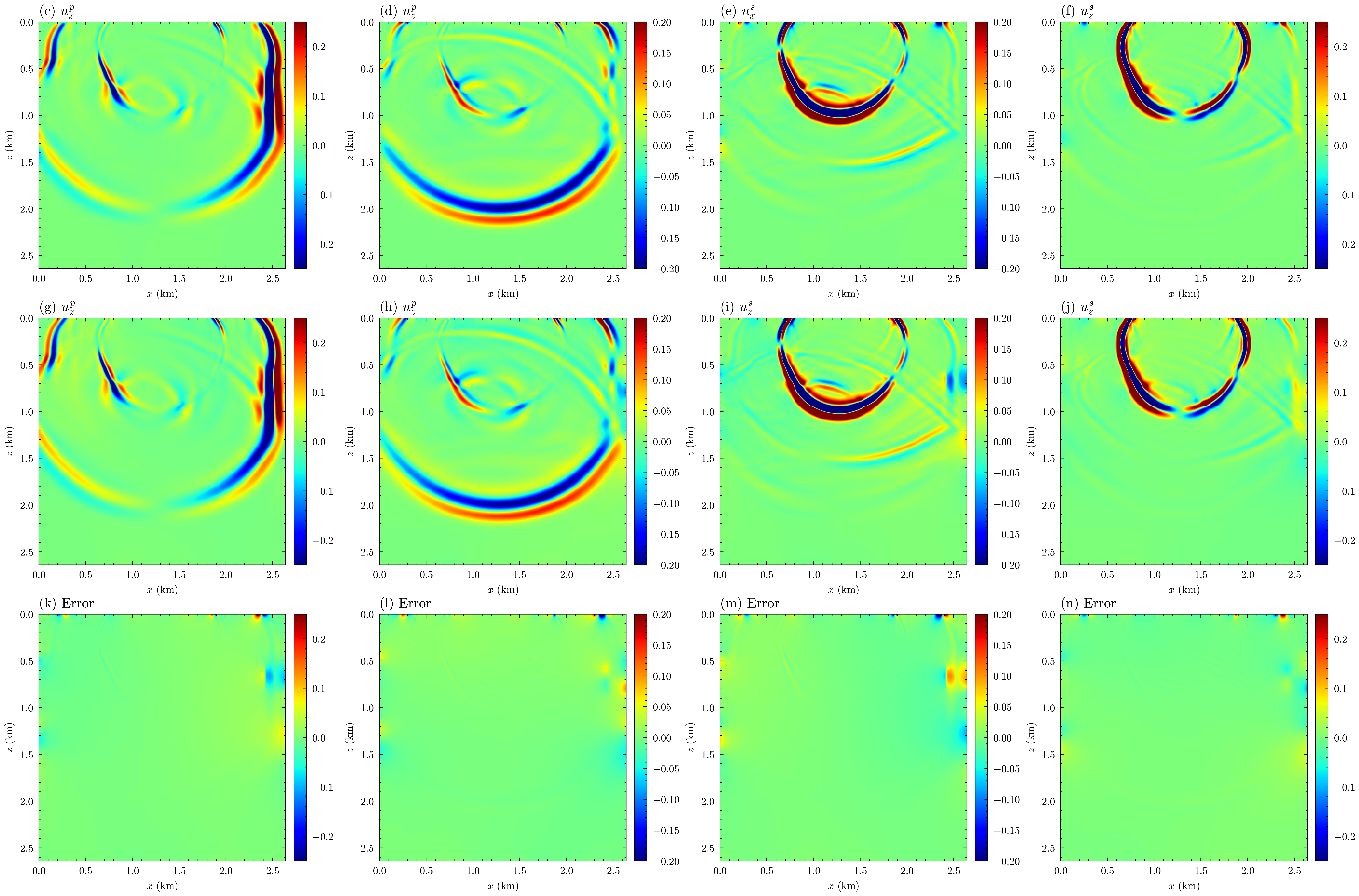}
    \caption{The same as Fig.\ \ref{fig:u_dec_homo_iso_comparation}, but for a non-homogeneous elastic medium shown in Fig.\ \ref{fig:GDM_velocity_model}.}
    \label{fig:u_dec_GDM_comparation}
\end{figure*}


\section{Elastic Wavefield Separation}
\label{sec:elastic_sep}

The elastic wave propagation in a homogeneous and isotropic medium is described by~\cite{Main_1980}
\begin{equation}
    \rho \frac{\partial^2\mathbf{u}}{\partial t^2} = \mathbf{f} + (\lambda + 2\mu)\nabla(\nabla\cdot\mathbf{u}) - \mu \nabla\times\nabla\times\mathbf{u}\,, \label{eq:elastic_wave_eq}
\end{equation}
where $\mathbf{u}$ is the elastic wavefield, $\mathbf{f}$ is the source wave term, $\rho$ is the density, and $\lambda$ and $\mu$ are the Lamé coefficients. The longitudinal and transverse wave velocities are defined as $v_p = \sqrt{(\lambda + 2\mu)/\rho}$ and $v_s = \sqrt{\mu/\rho}$, respectively.

The elastic wavefield can be decomposed into P and S waves using the Helmholtz decomposition~\cite{Zhu}, that is,
\begin{equation}
    \mathbf{u} = \mathbf{u}^p + \mathbf{u}^s\,,
\end{equation}
where these fields satisfy $\nabla\times\mathbf{u}^p = 0$ and $\nabla\cdot\mathbf{u}^s = 0$. Accordingly, these modes can be obtained through the relations
\begin{equation}
    \mathbf{u}^p = \nabla(\nabla\cdot\mathbf{w})\,, \quad \mathbf{u}^s = -\nabla\times\nabla\times\mathbf{w}\,,
\end{equation}
under the assumption that the auxiliary vector field $\mathbf{w}$ satisfies the following vector Poisson equation:
\begin{equation}
    \nabla^2\mathbf{w} = \mathbf{u}\,. \label{eq:vector_poisson}
\end{equation}

The computational cost involved in solving Eq.\ \eqref{eq:vector_poisson} grows with the number of spatial dimensions, which makes it cumbersome to apply it to large-scale problems. It can be diminished by introducing a scalar Poisson equation~\cite{ZhengYao}:
\begin{equation}
    \nabla^2 w = \nabla\cdot\mathbf{u}\,,\label{eq:scalar_poisson}
\end{equation}
where the elastic modes are now recovered by computing:
\begin{equation}
    \mathbf{u}^p = \nabla w\,, \quad\mathbf{u}^s = \mathbf{u} - \mathbf{u}^p\,.\label{eq:P-S_modes_scalar}
\end{equation}

In contrast to traditional decomposition methods, which use a scalar field for P and a vector field for S, the fully vector formulations presented above yield elastic modes with physically meaningful interpretations, preserving both amplitude and phase information~\cite{Zhu}.

\section{Physics-Informed Neural Network}
\label{sec:pinn}

We consider a fully connected feed-forward neural network (see Fig.\ \ref{fig:pinn}) comprising $L+1$ layers, where $L-1$ of these layers are hidden. The first input layer corresponds to the spatial coordinates $\mathbf{x} = (x, z) \in \mathbb{R}^2$, while the last one is the output solution $w(\mathbf{x}) \in \mathbb{R}$. In this construction, there are $k_l$ neurons in the $l$-th hidden layer. The layers are connected through the weight vector $W_{ki}^l$ and bias elements $B_{k}^l$. Therefore, the output of the $k$-th neuron in the $l$-th layer is given by a weighted sum of the outputs from the preceding layer~\cite{haykin}
\begin{equation}
    w_k^l = \sigma\left(\sum_{i=1}^{k_l-1} W_{ki}^l w_i^{l-1} + B_k^l\right)\,, \label{eq:nn}
\end{equation}
where $\sigma$ is a nonlinear activation function.

To mitigate the spectral bias~\cite{pmlr-v97-rahaman19a} in predicting high-frequency information present in elastic wavefields, we include Fourier features~\cite{NEURIPS2020_55053683, WANG2021113938, song_fourier} in our PINN architecture. Therefore, the spatial coordinates are first transformed into a higher-dimensional representation by the mapping function $\gamma(\mathbf{x})\in \mathbb{R}^{2n}$, given by:
\begin{equation}
    \gamma(\mathbf{x}) = \left(\sin(2\pi\mathbf{B}^{\mathrm{T}}\mathbf{x}),   \cos(2\pi\mathbf{B}^{\mathrm{T}}\mathbf{x})\right)^\mathrm{T}\,, \label{eq:FF}
\end{equation}
where $\mathbf{B}\in\mathbb{R}^{n\times2}$ carries the frequencies used in the mapping. After this embedding step, the transformed input $\gamma(\mathbf{x})$ combined with the original coordinates $\mathbf{x}$ is fed as input into the network (see Fig.\ \ref{fig:pinn}). We emphasize that incorporating Fourier features does not increase the training cost, since this mapping function does not contain any trainable parameters.

Finally, this Fourier feature-embedded PINN finds an approximated solution of Eq.\ \eqref{eq:scalar_poisson} by identifying the set of weighting and bias parameters that minimize the loss function
\begin{equation}
    \mathcal{L} = \mathcal{L}_{\mathrm{BC}} + \mathcal{L}_{\mathrm{PDE}}\,, \label{eq:loss}
\end{equation}
with the boundary condition and PDE terms having the following form
\begin{align}
    \mathcal{L}_{\mathrm{BC}} &= \frac{1}{N_b}\sum_{k=1}^{N_b} \left|w_k\right|_2^2\,, \\
    \mathcal{L}_{\mathrm{PDE}} &= \frac{1}{N_c}\sum_{\ell=1}^{N_c} \left|(\nabla^2 w - \nabla\cdot\mathbf{u})_\ell\right|_2^2\,,
\end{align}
where $N_c$ and $N_b$ are the numbers of collocation and boundary points used for the training, respectively.

\section{Results and Discussion}
\label{sec:results}

In this section, we employ physics-informed neural networks (PINNs) to address the elastic separation problem in both homogeneous and non-homogeneous elastic models. In each case, we compare our results with those obtained from a well-established numerical method based on the discrete sine transform (DST)~\cite{DST}. The solution of the elastic wave equation, Eq.\ \eqref{eq:elastic_wave_eq}, is obtained using the finite-difference method.

\subsection{Homogeneous Case}

We consider a homogeneous model with elastic parameters $\rho = 0.44 $ g/cm$^3$, $v_p = 4.0 $ km/s, and $v_s = 2.35 $ km/s. The spatial domain consists of an area of 1 km $\times$ 1 km, and it is discretized using a uniformly spaced grid consisting of 401 $\times$ 401 collocation and 1608 boundary points. A Ricker wavelet with a central frequency of 10 Hz was used as the vertical source at the center of the model. 

The network was built with eight hidden layers, which gradually decrease as follows: 256, 256, 128, 128, 64, 64, 32, 32. We used a sine function as the activation function in Eq.\ \eqref{eq:nn}, which is well-suited for describing oscillatory phenomena~\cite{song2022, Buzaev2024}. Moreover, in Eq.\ \eqref{eq:FF} we used $n = 4$ frequencies randomly sampled from a Gaussian distribution with zero mean and unit variance~\cite{NEURIPS2020_55053683}. In the training processes, we used the AdamW optimizer with an initial learning rate of $0.001$. The network was trained for $1.2\times 10^4$ epochs, with the learning rate reduced by half every 2000 epochs (see the loss curve in Fig.\ \ref{fig:Total_loss_vp}). 

The horizontal and vertical components of the elastic wavefield captured at $t = 0.25$ s are shown in Fig.\ \ref{fig:u_dec_homo_iso_comparation}(a) and Fig.\ \ref{fig:u_dec_homo_iso_comparation}(b), respectively. The conventional numerical and our proposed elastic separation method are shown in the second and third rows of Fig.\ \ref{fig:u_dec_homo_iso_comparation}, respectively. The difference between them is nearly zero, as shown in the fourth row of Fig.\ \ref{fig:u_dec_homo_iso_comparation}. Using this numerical method as reference, we obtain a global MSE error of $1.79\times10^{-6}$. Minimal artifacts occur at the boundary domain, which can be attributed to the difficulty PINN has in representing sharp variations. Despite this observation, these results indicate that our method can accurately separate elastic modes in homogeneous media. 

After training, the averaged execution time to separate the elastic modes with PINN is $9.36 \pm 0.37$ ms, while the conventional one is $0.61 \pm 0.07$ ms. This indicates that the proposed PINN method is approximately an order of magnitude slower than the numerical solver, which utilizes an optimized fast Fourier transform routine to solve the problem in the wavenumber domain. However, we argue that this additional computational cost can be compensated by PINN's flexibility in accommodating irregular geometries and noisy data, without requiring substantial modifications to its formulation. 

Remarkably, our results are similar to those obtained in Ref. \cite{mu2025separationpinnphysicsinformedneuralnetworks}, where two independent networks are employed to predict the horizontal and vertical components of the elastic wavefield. In contrast, our method uses a single network, which significantly reduces computational costs and makes the proposed approach more suitable for large-scale data scenarios.




\subsection{Non-homogeneous Case}

We now investigate the capabilities of wavefield separation in a more realistic Earth model. To this end, we consider a subsurface structure whose P-wave velocity profile is illustrated in Fig.\ \ref{fig:GDM_velocity_model}. From the top down, its geological structure consists of a water layer (up to $\sim 0.1$ km of depth), followed by a post-salt marine shale, a variable thickness salt body (see details in Ref. \cite{gm0}). The transverse velocity $v_s$ was approximated by $v_s = v_p/\sqrt{3}$. Accordingly, the density was estimated following Gardner's empirical relation $\rho = 0.31\times v_p^{1/4}$ (see Ref. \cite{Gardner}). Geological models featuring this salt structure exhibit significant P/S mode conversion due to the contrast in material properties at the interface sediment/salt~\cite{salt1_1997, salt2_2003, salt3_2013}. The spatial domain is discretized with 265 $\times$ 265 regularly spaced collocation points with 1064 along the boundary. To simulate a marine seismic acquisition, a 15 Hz Ricker wavelet was used as a vertical source at a depth of $150$ m.

We employ a deeper neural network in this case due to the velocity model complexity present in Fig.\ \ref{fig:GDM_velocity_model}. In this regard, we use ten hidden layers with the number of neurons per layer decreasing as follows: 512, 512, 256, 256, 128, 128, 64, 64, 32, 32. The network was trained using $3.5\times 10^4$ epochs, where the learning rate was decreased by half every $2.0\times 10^3$ epochs (see the loss curve in Fig.\ \ref{fig:Total_loss_gdm}). Moreover, we use the same number of Fourier frequencies as in the homogeneous case. Even with a velocity model that is three times as large and an expanded PINN architecture, the difference in execution time between the PINN and the conventional numerical solver remains of the same order of magnitude as in the first example.

The mode decomposition results are shown in Fig.\ \ref{fig:u_dec_GDM_comparation}. Similarly to the homogeneous case, the proposed method produces separated modes that are very similar to those obtained using the numerical method. The global MSE error is $4.94\times10^{-4}$ in this case. All complex scattering events occurring in this non-homogeneous medium are present. Besides the low-energy refracted and high-energy reflected waves, mode conversion phenomena can be identified, occurring mainly at the edge of the salt dome ($\sim 0.7$ of the depth and a horizontal distance of $\sim 0.6$ km), where there is a high velocity contrast. 

Interestingly, we observe that the difference between the two methods lies again at the boundary of the physical domain. PINN exhibits the same limitations as the homogeneous case in representing wavefronts that interact with domain boundaries. However, the numerical leakage effects related to residual S-wave contamination in the P-mode are substantially suppressed using the proposed approach. 

\section{Conclusion}
\label{conc}

We have addressed the elastic separation problem using a physics-informed neural network. Instead of following the traditional vector Helmholtz decomposition, we have adopted an alternative approach in which the elastic modes are obtained by solving an auxiliary scalar Poisson equation. We have significantly reduced network complexity compared to existing methods in the literature, as our approach enables the construction of a single neural network with a single output. Therefore, we leveraged the PINN’s ability to incorporate physical equations while exploiting mathematical relationships between vector operators, resulting in a dimensionally scalable machine learning method that does not require large amounts of training data. We have compared our proposed method with a well-established numerical approach, demonstrating that it can accurately separate elastic modes in complex media. We believe that this same approach is directly transferable to other scenarios where mode conversion of elastic waves in solids is relevant.

\begin{acknowledgments}

E. A. B. Alves, P. D. S. de Lima, M. S. Ferreira, J. M. de Araújo and C. G. Bezerra acknowledge Petrobras' support through the “Development of Seismic Inversion Methodologies for 4D Reservoir Monitoring” project at Universidade Federal do Rio Grande do Norte (UFRN), as well as the strategic importance of the support provided by ANP through the R\&D levy regulation. J. M. de Araújo and C. G. Bezerra acknowledge support from CNPq (Grant Nos. 311589/2021-9 and 310919/2025-8, respectively). The authors also thank the High-Performance Computing Center (NPAD) at UFRN for providing the computational resources that made this work possible.

\end{acknowledgments}

\bibliography{separation_pinn}

\end{document}